\documentclass[%
 aip,
 jmp,%
 amsmath,amssymb,
 reprint,%
]{revtex4-1}

\usepackage{graphicx}
\usepackage{dcolumn}
\usepackage{bm}

\def\gev{GeV/$c^2$}

\begin{document}


\title{Toward Single Electron Resolution Phonon Mediated Ionization Detectors}


\author{Nader Mirabolfathi, H. Rusty Harris, Rupak Mahapatra, Kyle Sundqvist, Andrew Jastram}
\affiliation{Department of Physics and Astronomy, Texas A\&M University}

\author{Bruno Serfass, Dana Faiez, Bernard Sadoulet}
\affiliation{Department of Physics, University of California at Berkeley}


\date{\today}

\begin{abstract}

Experiments seeking to detect rare event interactions  such as dark matter or coherent elastic neutrino nucleus scattering are striving for large mass detectors with very low detection threshold. Using Neganov-Luke phonon amplification effect, the Cryogenic Dark Matter Search (CDMS) experiment is reaching unprecedented RMS resolutions of ~ 14 eV$_{ee}$ . CDMSlite is currently the most sensitive experiment to WIMPs of mass $\sim$5 GeV/c$^{2}$ but is limited in achieving higher phonon gains due to an early onset of leakage current into Ge crystals. The contact interface geometry is particularly weak for blocking hole injection from the metal, and thus a new design is demonstrated that allows high voltage bias via vacuum separated electrode. With an increased bias voltage and a $\times$ 2 Luke phonon gain, world best RMS resolution of sigma $\sim$7 eV$_{ee}$ for 0.25 kg (d=75 mm, h=1 cm) Ge detectors was achieved. Since the leakage current is a function of the field and the phonon gain is a function of the applied voltage, appropriately robust interface blocking material combined with thicker substrate (25 mm) will reach a resolution of $\sim$2.8 eV$_{ee}$. In order to achieve better resolution of $\sim$ eV, we are investigating a layer of insulator between the phonon readout surface and the semiconductor crystals.
\end{abstract}


\maketitle

\section{Motivation: Ultra low threshold detectors for Dark Matter and Coherent Neutrino scattering detection}

A large body of astrophysical observations point to the fact that 85\% of the matter in the universe is not made of known standard model particles ~\cite{Ade:2013zuv}. Understanding the nature of this dark matter is of fundamental importance to cosmology, astrophysics, and high energy particle physics. Although Weakly Interacting Massive Particles (WIMPs) of the mass 10-100 \gev ~have been the main interest of the majority of direct dark matter detection experiments, recent claims for signal, together with compelling new theoretical models, are shifting the old paradigm toward broader regions in the dark matter parameter space well below 10 \gev \cite{CF1}. 

Very low energy recoils and small interaction rates from these low mass WIMPs are expected, thus large mass detectors with very low threshold are highly desired. These very low threshold detectors are \textit{sine qua non} for any attempt to detect very light mass ($<$1 \gev ) dark matter. They are also a necessary requisite for observing coherent elastic neutrino$-$nucleus scattering~\cite{CNS}, a standard model process that has recently been proposed as a sensitive and flavor invariant probe for sterile neutrinos, neutrino magnetic moment and other HEP physics~\cite{MItheory,CNS_Sterile}.

P. Luke had suggested to utilize very low noise readout designed for phonon mediated detectors to indirectly measuring ionization in semiconductor detectors~\cite{Luke_CDMSlite}. The measurement principle is based on the fact that carriers drifting through crystals under an applied electric field release phonons whose total energy is proportional to the interaction energy as well as the applied bias voltage:
\begin{eqnarray}
E_{Luke}=V_{bias}E/\epsilon
\label{eq:one}.
\end{eqnarray}
 Where $E$ is the energy of the interaction and $\epsilon$ is the average energy necessary to produce electron and hole pairs. Since the total signal is proportional to the bias potential, in the absence of any leakage current, the Signal-to-Noise Ratio (SNR) improves proportionally to the bias and can be improved down to single electron-hole sensitivity. CDMSlite is using this very sensitive method to search for low mass dark matter and is currently the most sensitive experiment for WIMPs of masses $<$ 5 GeV$/c^{2}$ \cite{Agnese:2013lua}. However, current generation CDMSlite detectors exhibit a leakage current for fields as low as 24 Volts/cm, thus single ionization sensitivity has not yet been realized. This early onset of leakage is in contrast with results from standard 77~K depleted Ge detectors that are usually operated with much larger fields ($\sim$1000 Volts/cm). Recent understanding of the CDMSlite interface shows that it is comprised of polycrystalline grain boundaries that allow charge leakage (to be published in parallel to this report). Here we report on our recent studies and success toward understanding this early onset of leakage current and our suggestions to improve ionization contacts for the ultimate single electron resolution detectors.

\section{Bulk breakdown versus carrier injection through contact}
We can think of three sources of excessive leakage current in CDMS detectors: Crystal break down, carrier injection through metal/germanium contact or conduction on non passivated free surfaces of detectors. The surface conduction is eliminated outright because of precise, clean crystal fabrication and handling. Furthermore, the amount of current leakage observed cannot be accounted for in surface current density without significant damage to the detector.

\subsection{Ge Crystal bulk breakdown}
The first evidence in eliminating bulk crystal breakdown is the fact that such breakdown is catastrophic and irreversible. We observe that reverse biased detectors that have previously experienced significant leakage do not demonstrate high leakage when appropriately biased. The second set of evidence is in the low electric field $\sim$30 Volts/cm at which high leakage is observed.

\begin{figure}
\includegraphics[width=1.5 in]{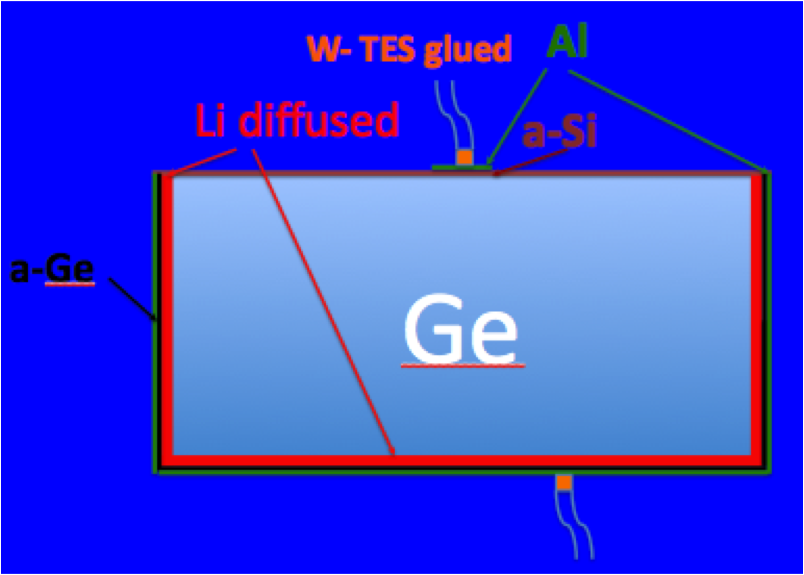}
\includegraphics[width=1.35 in]{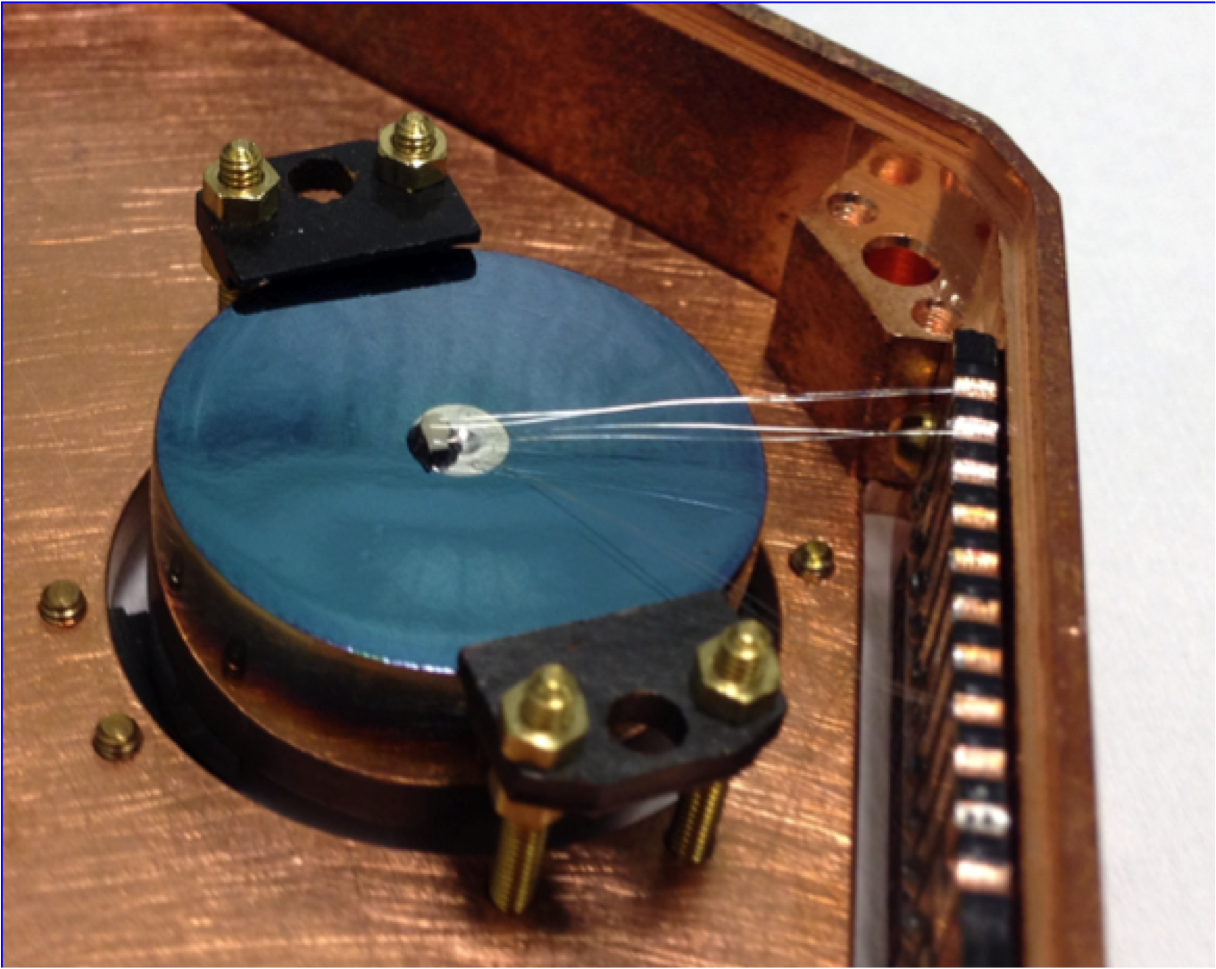}\\
\includegraphics[width=2.6 in]{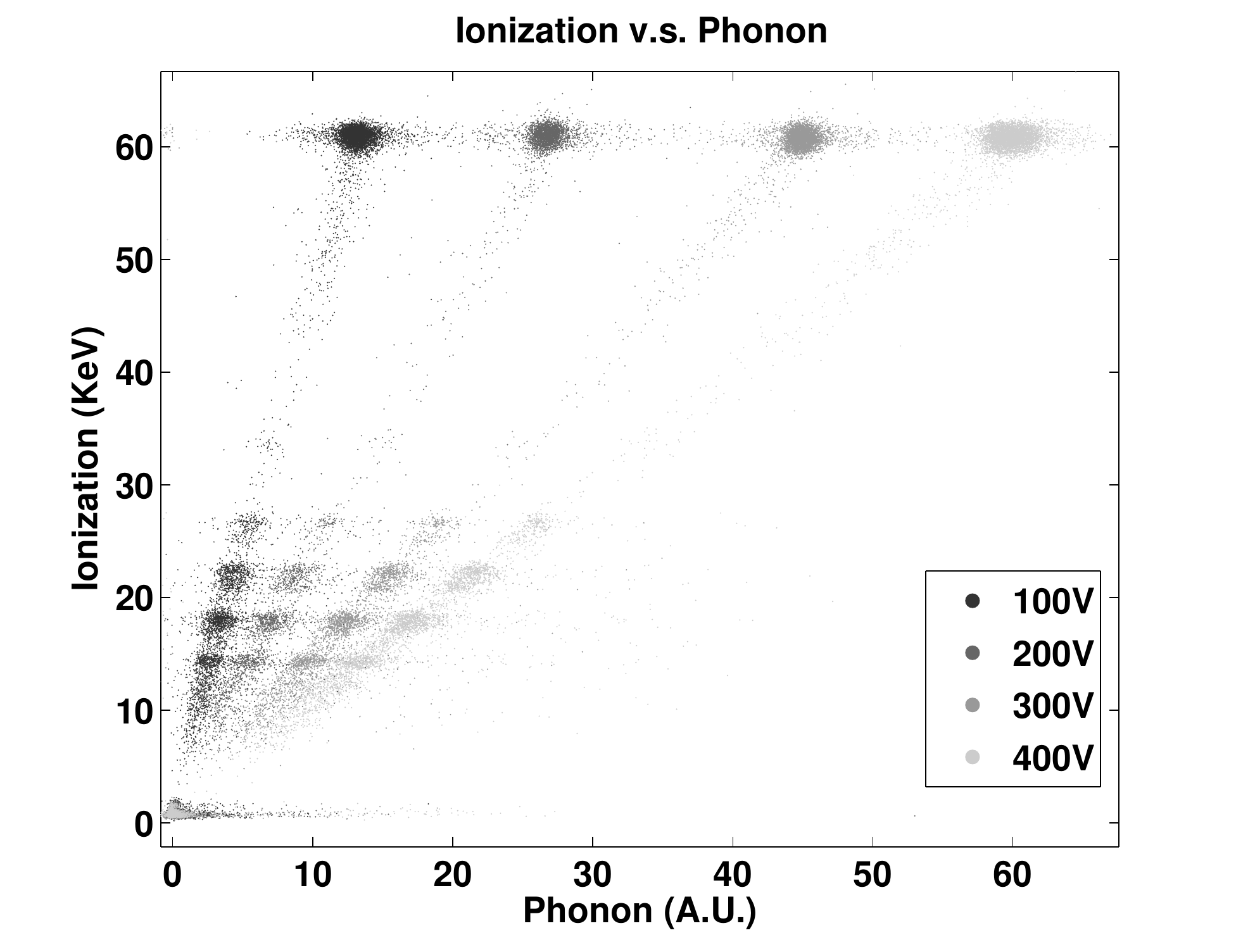}
\caption{\footnotesize(Top Left) Schematic of the PPC detector used in this work. The point contact is defined by a 3\,mm diameter vapor-deposited aluminum disk on the amorphous silicon passivation layer. The face opposite of the point contact and the cylindrical surface have lithium diffused to a depth of $\sim$1\,mm coated with sputtered amorphous Ge followed by vapor-deposited aluminum. (Top right) Photo of the detector in its copper housing. The long wire bonds connect the point contact and the TES thermistors to the CDMS Digital Interface Board and corresponding cold electronics. (Bottom) Ionization versus phonon measurement for various bias voltages up to 400 Volts. The phonon signal increase proportional to bias while ionization signal remain constant as expected.}
\label{fig:PPC}
\end{figure}

In a previous study we ruled out the breakdown in  high purity Ge crystals by operating a Majorana prototype P-type Point Contact Ultra pure Ge detcetor (PPC) in the similar setup wherein we operate CDMS detectors~\cite{LTD13_contactfree}. A 17 g PPC prototype was equipped with a tungsten Transition Edge Sensor (TES) thermistor and both ionization and phonon measured up to 400 Volts as shown in Fig.~\ref{fig:PPC}. There were no sign of crystal breakdown despite large fields present in the vicinity of the point contact (up to 7000 Volts/cm). This clearly ruled out the crystal breakdown. We also verified that the phonons gain is a linear function of applied voltage up to our limited DC voltage supply circuit.  

CDMS iZIPs interface structure is symmetric (amorphous$-$Si only), in contrast with LBNL PPC design wherein the positive and negative bias electrodes are interfaced with different material (amorphous$-$Si or amorphous$-$Ge) to the bulk (Figure~\ref{fig:PPC} Top left). A previous study by the LBNL group shows that the blocking properties of amorphous Si or Ge are different for different types of carriers and that amorphous$-$Ge better blocks holes than amorphous$-$Si~\cite{Amman2007}. However, these detectors are operated and tested at $>$77K, well above freeze-out of Ge and high voltage operations at the low temperatures needed for phonon sensing may require different contacts.



\begin{figure}
\includegraphics[width=1.2 in]{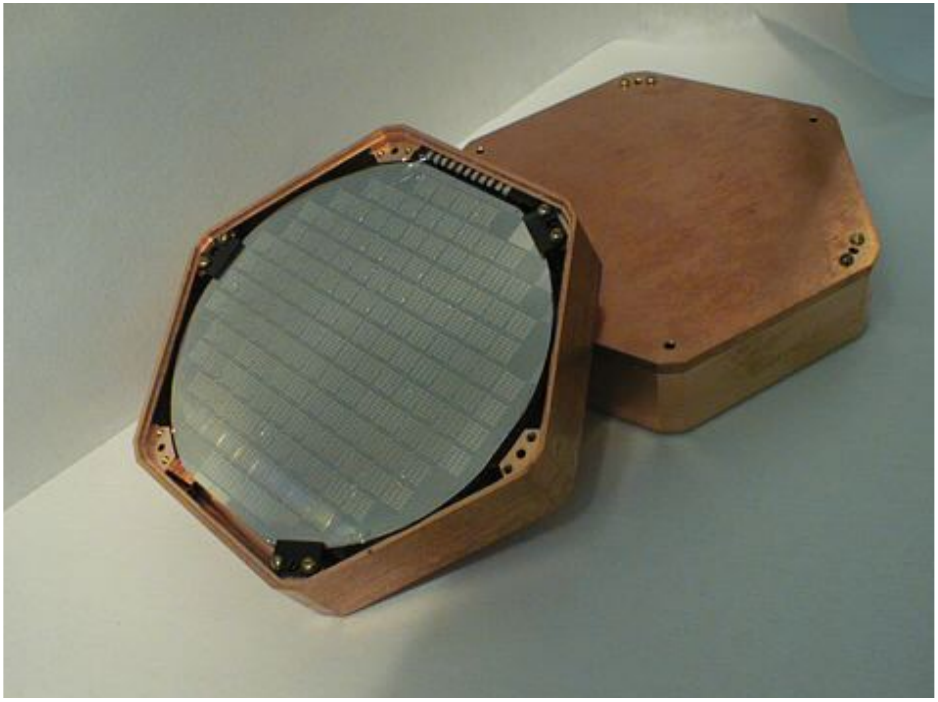}
\includegraphics[width=1. in]{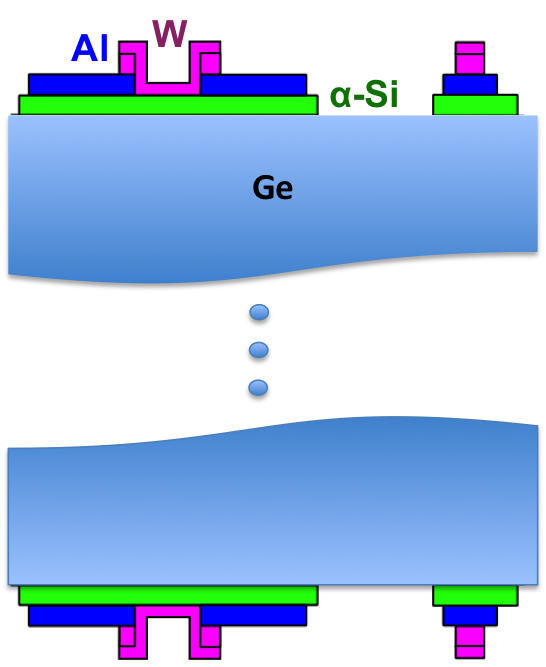}
\includegraphics[width=1. in]{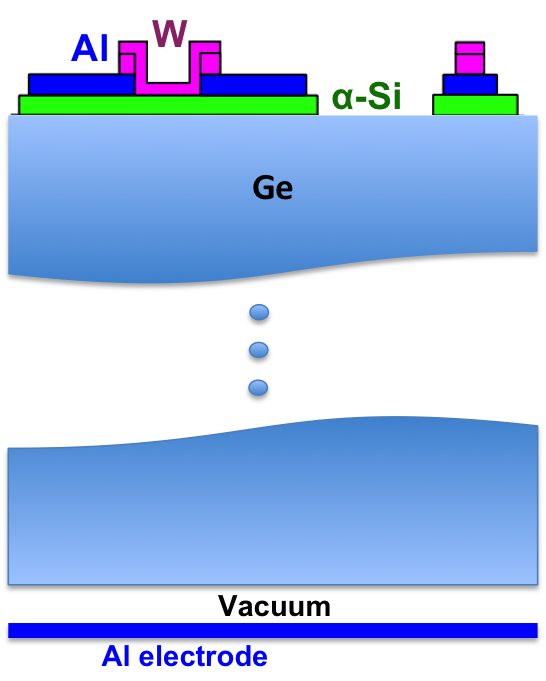}
\caption{\footnotesize(Left) CDMS ZIP phonon readout consist of 1024 W TES athermal phonon sensors covering one face of $\varphi=$75 mm h=10 mm Ge crystal. (Middle) CDMS detector contact geometry is symmetric on both faces. The aluminum bias contacts are interfaced through a layer of 50 nm $\alpha$-Si to the Ge crystal (Right) One face of the detector studied in this work is not processed (bare polished Ge) and the crystal is biased via a $\sim$500 micron gap.}
\label{fig:Contactfree}
\end{figure}

\subsection{New Contact Geometry}

To break CDMS design symmetry, we fabricated a detector wherein, we eliminate one interface and bias the detector with a flat aluminum electrode separated from the crystal via a small gap  $\sim$500 $\mu$m (Fig.~\ref{fig:Contactfree}). One face of a 0.250 kg Ge substrate was processed with athermal phonon sensors similar to CDMS ZIP detectors and was covered with athermal phonon tungsten (W) Transition Edge Sensors (TES). The phonon sensors are also acting as the ground for biasing purposes and are interfaced to the detector through a layer of $\alpha$-Si. the other face of the detectors was left with bare polished Ge. The detector is biased via a gap of $\sim$500 $\mu$m. The bias electrode was beveled to avoid the high fields associated with the sharp features at the edges of the electrodes. Since no carrier injection is expected from the contact-free face of the detector, this new biasing scheme allows for independent study of the Al/$\alpha$-Si/Ge interface properties.

The detector was mounted with a collimated $^{241}Am$ source on the phonon face of the detector and was installed in Berkeley $^{3}$He-$^{4}$He dilution refrigerator and cooled to $<$0.04 K. We discuss results from high voltage operation of this detector.  A Light Emitting Diode (LED) with photon energies slightly above the Ge band gap was mounted close to the detector volume. By grounding the detector and pulsing the LED, large number of electrons and holes are generated to neutralize and reset the detector. Since the charge carriers that arrive to the bare face of the Ge has no path to leave the crystal, this LED pulsing is crucial in order to eliminate the accumulated charge and the counter electric field thereof. 

\section{Experimental results and discussion}

Leakage current in the new contact-free geometry for both positive and negative bias polarities was measured. The phonon surface is always held at the ground potential. All the bias polarities mentioned in this paper are referenced to the phonon face electrode. Therefore, for the positive bias polarities the holes drift toward the phonon surface and electrons toward the bias electrode and vice versa. Since there is only one contact present in this detector design (phonon surface), we can study the polarity dependence of current leakage through Al/pc-Si/Ge with this device. 

We use phonon noise performance as a metric to evaluate leakage. The carriers that are randomly leaking through the Ge electrode interface will also drift within the crystal volume producing Neganov-Luke phonons and will appear as an increase in phonon readout noise. We found this method of leakage current measurement much more sensitive than the standard ionization direct charge readout.   

\begin{figure}
\begin{center}
\includegraphics[width=3 in]{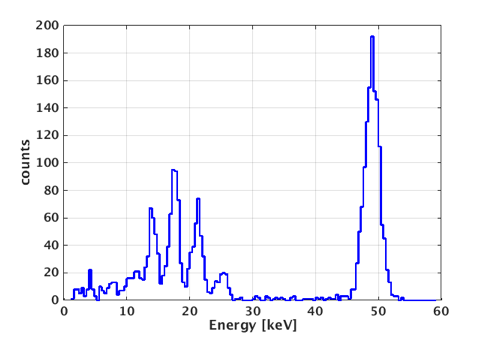}
\includegraphics[width=3 in]{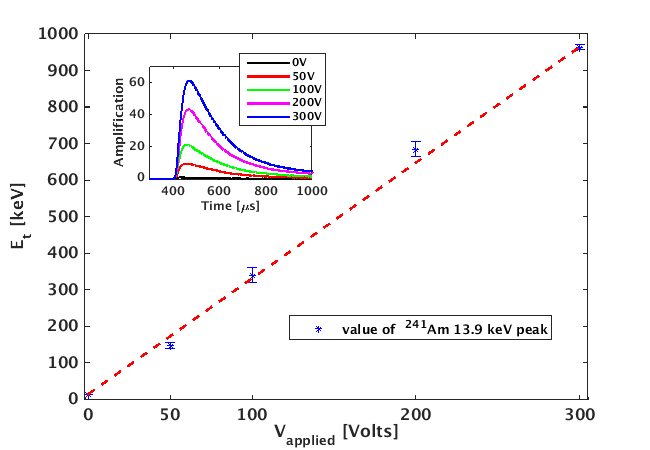}
\caption{\footnotesize Top) Spectrum of $^{241}$Am source obtained at V$_{bias}$=300 Volts. The 60 keV photon line appears at lower energy due to large phonon gain and TES saturation. Bottom) Linear phonon gain and phonon pulses for 13.9 keV line as a function of bias. The pulse amplitudes increase linearly with bias as expected. We can use the line fit to estimate the vacuum gap between the bias electrode and the crystal.}
\label{fig:Luke linearity}
\end{center}
\vspace{-20pt}
\end{figure}

In clear agreement with the previous studies with PPC detectors, we observed a leakage asymmetry with respect to the bias polarity in this device. The detector sustains up to 350 Volts in positive polarity but exhibits significant leakage for negative polarity even for biases as low as 20 Volts shown in Fig. \ref{fig:Luke linearity}. In agreement with LBNL previous results, this suggests that the amorphous$-$Si interface is a much weaker blocking layer for hole injection from Al compared to electron injection from the same.  

Since the detector is biased via a gap of $\sim$500~$\mu$m, the actual potential across the detector (V$_{Ge}$) is smaller than the applied voltage (V$_{applied}$) by the ratio of detector capacitance (100 pF) and the gap. We use the phonon amplification gain for $^{241}$Am 13.9 keV line compared to V=0 Volts data to estimate the true voltage across the detector. We found the V$_{Ge}$/V$_{applied}$ ratio to be $\sim$0.7. We also confirmed Neganov-Luke gain linearity with respect to applied voltages up to 350 Volts (corresponding to V$_{Ge}$=250 Volts) as shown in Fig.~\ref{fig:Luke linearity}. We observed slight noise increase associated with the rise of leakage current in the detector for V$_{Ge}$$\sim$ 140 Volts across the crystal. Assuming an almost uniform electric field in this geometry, this corresponds to an almost uniform electric field of $\sim$140 Volts/cm within the crystal volume.   

\begin{figure}
\begin{center}
\includegraphics[width=2.8 in]{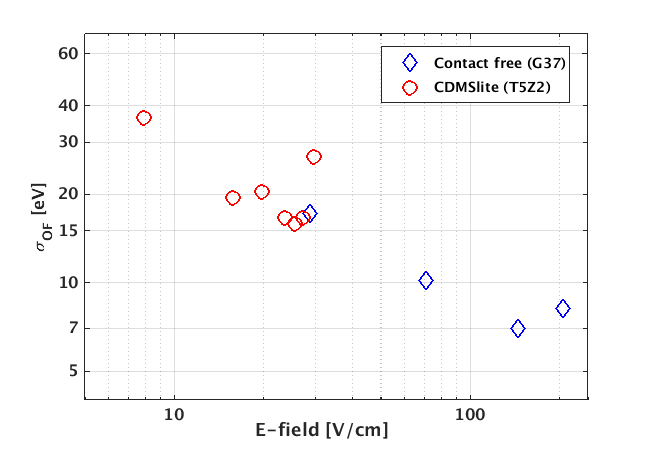}
\includegraphics[width=2.8 in]{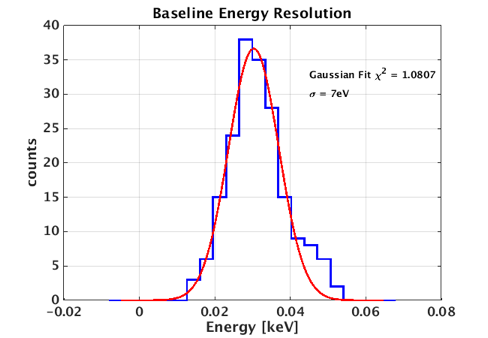}
\caption{\footnotesize Top) Ionization base line RMS resolution as a function of E-field for CDMSlite~\cite{Rito2015} (red circles) and contact-free detector reported in this work. Clear sign of leakage is observed for fields above 24 Volts/cm in CDMSlite detectors. The contact-free detector RMS resolution continues to improve for E$>$100 Volts/cm. Bottom) Best RMS resolution measured in this geometry is slightly below 7 eV. The phonon gain at constant electric field scales linearly with thickness, thus we expect to have $<$ 3 eV resolution using CDMSlite detector geometry.}
\label{fig:resolution}
\end{center}
\vspace{-20pt}
\end{figure}

RMS resolution was measured for every applied bias voltage by fitting phonon templates generated from $^{241}$Am 13.9 keV events to the randomly triggered baseline traces. We fit a gaussian function to the resulted distribution and calculate the $\sigma$ of the distribution. In Fig. \ref{fig:resolution} we compare phonon resolution progression with respect to the applied electric field between CDMSlite and contact-free geometries.  As expected the resolution improves linearly with bias up until the detector exhibits significant leakage, thus increased phonon noise. The leakage onset is clearly occurs at much smaller fields in CDMSlite ($\sim$25 Volts/cm) than contact free. The best RMS resolution obtained with this contact-free prototype detector corresponds to V$_{Ge}$$\sim$140 Volts  was slightly below 7 eV. This is the world best ionization baseline resolution for a detector of this size. 

The prototype contact-free detector presented here is a 1 cm i.e. $\times$2.5 thiner than CDMSlite detectors. Since the leakage current is a function of the electric field and the phonon gain is a function of potential, we expect a straightforward improvement of $\times$2.5 in phonon resolution down to $\sigma$=2.8 eV using the contact-free biasing scheme on 2.5 cm thick CDMSlite geometry. Our detectors are sensitive to athermal phonons, thus the larger volume of the detector should not affect phonon collection efficiency since majority of the phonons will relax in the metallic sensors. 

In order to reach better resolutions, the interface device physics must be adapted to maintain low hole and electron injection from the Al while allowing phonon transfer to the TES. We propose that a dielectric with significant electron and hole band edge offsets, but sufficiently thin for phonon transparency is the best option. However, formation of quality dielectrics at low temperatures so as to not create di-vacancies and other charge trapping defects is a critical challenge. We are currently investigating SiO$_{2}$, sapphire and nitrides for just such an application.

\section{Conclusion}
Combined results form prior tests of LBNL PPC and our new contact-free detectors point to the specific CDMSlite contact design to be the source for low field leakage current observed in CDMS. We found that Al/$\alpha$-Si/Ge has asymmetric blocking properties for electron versus hole injection and provides better electron blockage. Our results are in agreement with previous LBNL results wherein amorphous Ge had beed suggested to block hole injection. CDMS is currently studying $\alpha$-Ge  layers as an alternative for contact interface. 

Our new contact-free detector can sustain substantially larger bias than CDMSlite detectors. Using Neganov-Luke phonon amplification method, we achieved world best RMS resolution of 7 eV with this prototype detector. Further $\times$2.5 gain in resolution (to $\sigma$=2.8 eV) is readily available by scaling the detector thickness to that of CDMSlite detectors. However, in order to reach resolutions as low as single electrons, we will need to fully isolate the phonon readout surface of the detector. One possibility is to use a thin layer of SiO$_{2}$ to interface the phonon readout surface to the detector. It has been recently suggested that operating detectors with lower T$_{c}$ phonon sensors will 
improve resolution to the eV scale ~\cite{Pyle2015}. The combination of this work and those studies can lead to the ultimate single electron resolution detectors with event$-$by$-$event discrimination capability. 

\section{Acknowledgments}
This work has been supported in part by the National Science Foundation under the awards PHY$-$1102841 and PHY$-$1242645 and by the University of California MRPI Search For Dark Matter Initiative and DOE grant DE-SC0004022. The authors wish to acknowledge CDMS collaboration support and very useful discussions with Paul Luke, Kai Vetter, Mark Amman, Ryan Martin, Paul Brink, Matt Pyle, Ritto Thakur. This work wouldn't be accomplished without technical support form Mark Platt, James Phillips and Matthew Cherry.


\end{document}